\documentclass[aps,prd,twocolumn,showpacs]{revtex4}
\usepackage{graphicx}
\usepackage{epsfig}
\usepackage{dcolumn}
\usepackage{amsmath}
\usepackage{bm}

\begin{document}

\title{Five-dimensional PPN formalism and experimental test of Kaluza-Klein theory}

\author{Peng Xu\footnote{Email: moooonbird@gmail.com} and Yongge Ma\footnote{Email: mayg@bnu.edu.cn}}
\affiliation{Department of Physics, Beijing Normal University,
Beijing 100875, China}

\begin{abstract}
The parametrized post Newtonian formalism for 5-dimensional metric
theories with a compact extra dimension is developed. The relation
of the 5-dimensional and 4-dimensional formulations is then
analyzed, in order to compare the higher dimensional theories of
gravity with experiments. It turns out that the value of post
Newtonian parameter $\gamma$ in the reduced 5-dimensional
Kaluza-Klein theory is two times smaller than that in 4-dimensional
general relativity. The departure is due to the existence of an extra 
dimension in the Kaluza-Klein theory. Thus the confrontation between the reduced
4-dimensional formalism and Solar system experiments raises a severe
challenge to the classical Kaluza-Klein theory.

\end{abstract}

\pacs{04.50.+h, 04.25.Nx, 04.80.Cc}

\maketitle

As a candidate of fundamental theory, Kaluza-Klein (KK) theory
unifies gravity with electromagnetic field (or Yang-Mills field) by
certain higher dimensional general relativity (GR) \cite{kk}
\cite{bla}. Since the original 5-dimensional (5D) KK theory was
proposed by Kaluza \cite{kk2} and Klein \cite{kk1}, considerable
works have been done along this line \cite{duff} \cite{kk3}
\cite{wesson}. The fantastic idea that our spacetime has extra
dimension, promotes various higher dimensional theories, including
the well-known string theory \cite{str}. Besides the potential
function to unify the fundamental interactions, higher dimensional
gravity theories are also shown to be effective in accounting for
the dark constituent of the universe (see e.g. \cite{qq}). Given the
fascinating virtues of extra dimensions, it becomes very desirable
to confront higher dimensional theories of gravity with experiments.
Works on this subject can be traced back to 1980's \cite{pe}
\cite{Div}, while no agreement has been obtained in the literature.
Different classes of solutions to higher dimensional GR are designed
to represent Solar system (for soliton-like solution see \cite{we}
\cite{we2} \cite{liu2000}, for Schwarzschild-like solution see
\cite{pe2} \cite{rah}). However, whether the available experimental
data permit higher dimensional theories gets quite different answers
in different approaches. These ambiguities are caused by the freedom
in choosing higher dimensional solutions which are supposed to
represent the Solar system in 4 dimensions. On the other hand, in
4-dimensional (4D) case, a general framework, called Canonical
Parameterized Post-Newtonian (PPN) Formalism, was established by
Nordtvedt, Will et al. \cite{nor} \cite{will1} \cite{will2} in 1970s
as a basic tool to connect gravitational theories with the Solar
system experiments. In PPN formalism, the perturbative metric of a
gravitational theory, which is generated by the matter distribution
of the Solar system, is expanded by orders in terms of linear
combinations of post Newtonian potentials. The differences among
various metric theories are represented by the coefficients (the PPN
parameters) of these post Newtonian potentials. Because of its high
accuracy and well-defined procedure, PPN formalism has attained
great achievements in testing 4D metric theories by Solar system
experiments \cite{wilbook} \cite{willliving}. Thus, some crucial
issues arise naturally. Is there a higher-dimensional PPN formalism?
If there is, what is the relation between the higher dimensional
formalism and the 4D one? More crucially, can one test higher
dimensional theories by the accurate Solar system experiments
without the ambiguities motioned above? The purpose of this letter
is to address these issues first in terms of 5D gravity theories
with a compact extra dimension. A 5D PPN formalism will be
developed. Its relation with the 4D formalism will be set up. As one
will see without any ambiguities, the concrete analysis reveals a
severe contradiction between KK theory and the Solar system
experiments.

The $5$D gravitational theories which we consider are defined on
some 5-manifold with topology $\mathbf{M}^{4}\times\mathbf{S}^{1}$,
where $\mathbf{S}^{1}$ is a compact extra dimension of radii $R$.
Both gravity and matter fields are assumed to be distributed over
the 5-manifold.\ Similar to 4D PPN formalism, the post Newtonian
coordinates system is chosen as certain asymptotic (in 4D sense)
flat system $\{t,x^{m}\},m=1,2,3,5$, where $x^{5}$ is the coordinate
of extra space. Since the compactification radii $R$ is sufficiently
small, a killing vector field $\xi^\mu$ arises naturally along the
extra dimension in the low energy regime \cite{bla}. It is
convenient to take an adapted coordinate system such that its fifth
coordinate basis vector $(\frac{\partial}{\partial x^{5}})^{\mu}$
coincides with $\xi^\mu$. The 5-metric reads
$\widetilde{g}_{\mu\nu}=\widetilde{\eta}_{\mu\nu}+\widetilde{h}_{\mu\nu
}$ with signature (-,+,+,+,+), where $\widetilde{h}_{\mu\nu}$ is the
perturbative metric generated by the matter distribution, e.g., the
Solar system. The gauge is chosen so that the spatial part of
$\widetilde{h}_{\mu \nu}$ is diagonal. As in Canonical PPN
Formalism, we will expand $\widetilde{h}_{\mu\nu}$ by orders in
terms of linear combinations of our generalized post Newtonian
potentials which are functionals of matter variables. We assume that
the matter composing the Solar system can be idealized as a perfect
fluid. The matter variables which we considered for the 5D perfect
fluid in Solar system include: 5D rest mass density
$\widetilde{\rho}$, 5D pressure $\widetilde{p}$ for the matter flow,
the ratio $\widetilde{\Pi}$ of 5D specific energy (including
compressional energy, radiation, thermal energy, etc.) density to 5D
rest mass density, and the coordinate velocity $\widetilde{v}^{m}$
of material particles or matter flow in post Newtonian frame. The
first three 5D matter variables give the corresponding
 effective 4D matter variables as%
\begin{equation}
\int\sqrt{\widetilde{g}_{55}}\widetilde{\rho}dx^{5}=\rho,\text{\
}\int \sqrt{\widetilde{g}_{55}}\widetilde{p}dx^{5}=p\text{,\ }\int
\sqrt{\widetilde{g}_{55}}\widetilde{\rho}\widetilde{\Pi}dx^{5}=\rho
\Pi.\label{55}%
\end{equation}
The general 5D post Newtonian potentials which\ we used for KK-like
theories are
$\widetilde{U},\widetilde{\Phi}_{1},\widetilde{\Phi}_{2},\widetilde{\Phi
}_{3},\widetilde{\Phi}_{4},$ and $\widetilde{V}_{m}$, which satisfy
respectively the 5D Poisson equations with respect to the flat
spatial background as:
\begin{align}
\nabla^{2}\widetilde{U}
=-\frac{16}{3}\pi\widetilde{G}\widetilde{\rho }, \ \
\nabla^{2}\widetilde{\Phi}_{1} =-\frac{16}{3}\pi\widetilde{G}
\widetilde{\rho}v^{2},  \nonumber \\
\nabla^{2}\widetilde{\Phi}_{2} =-\frac{16}{3}\pi\widetilde{G}
\widetilde{\rho}\widetilde{U},\ \ \nabla^{2}\widetilde{\Phi}_{3}
=-\frac{16}{3}\pi\widetilde{G} \widetilde{\rho}\widetilde{\Pi},
\nonumber  \\
\nabla^{2}\widetilde{\Phi}_{4}
=-\frac{16}{3}\pi\widetilde{G}\widetilde {p}, \ \
\nabla^{2}\widetilde{V}_{m}
=-\frac{16}{3}\pi\widetilde{G}\widetilde{\rho
}\widetilde{v}_{m},\nonumber
\end{align}
where $\widetilde{G}$\ denotes the 5D gravitational constant and we
use the unit where the velocity of light $c=1$. Note that one may
add more potentials in this framework in order to consider more
complicated 5D theories. Note also that the upper bound of the
compactification radii $R$ is constrained by the
tests of gravitational inverse-square law to be about $10^{-4}%
\operatorname{m}%
$ \cite{liu}, which is sufficiently small compared with the
characteristic
length $10^{12}%
\operatorname{m}%
$ of Solar system. With this condition we can estimate the order
relations of matter variables and potentials. Since
$|\widetilde{v}|\ll1$, we denote its order of smallness as
$\widetilde{v}\sim\mathcal{O}(1)$. Note that in the adapted
coordinate system the 5-metric components take the form \cite{kk2}
\cite{ma}:%
\begin{equation}
\widetilde{g}_{\mu\nu}=\left(
\begin{array}
[c]{cc}%
g_{\alpha\beta}+\phi B_{\alpha}B_{\beta} & \phi B_{\alpha}\\
\phi B_{\beta} & \phi
\end{array}
\right)  ,
\end{equation}
where $\alpha,\beta=0,1,2,3$. Thus, the "effective" 4-spacetime can
be understood as $(M^{4},g_{\alpha\beta})$ with the local coordinate
system $\{x^{\alpha}\}$ \cite{ma} \cite{yang}. Denote the 5-velocity
of a test particle as $\widetilde{U}^{\mu}$, then the 4-velocity of
the particle in $M^{4}$ is defined as \cite{ma}
\begin{equation}
U^{\alpha}=\frac{\widetilde{U}^{\alpha}}{\sqrt{-\widetilde{U}^{\alpha
}\widetilde{U}_{\alpha}}},\label{u4}%
\end{equation}
where $\widetilde{U}^{\alpha}\widetilde{U}_{\alpha}\equiv
g_{\alpha\beta }\widetilde{U}^{\alpha}\widetilde{U}^{\beta}$. From
Eq.(\ref{u4}) one can estimate the order relation between the
coordinate velocities in five and four dimensions as
$\widetilde{v}^{i}=v^{i}+\mathcal{O}(3)$.\ From Virial's theorem we
have $\widetilde{v}^{2}\sim\widetilde{U}\sim\mathcal{O}(2).$ Since
the scale of the extra dimension is very small, one can approximate
the solution of 5D Poisson equations by that of the corresponding 4D
equations. Hence the Newtonian gravitational potentials in five and
four dimensions are of the same order, i.e., $\widetilde{U}\sim U$.
The order relations between the matter variables in five and four
dimensions can be estimated from
Eq.(\ref{55}) as $\widetilde{\Pi}\sim\Pi$ and $\frac{\widetilde{p}}%
{\widetilde{\rho}}\sim\frac{p}{\rho}$. Therefore, in the light of
the order
relations in 4D PPN theory \cite{wilbook}, we obtain $\widetilde{U}%
\sim\widetilde{\Pi}\sim U\sim\Pi\sim\mathcal{O}(2)$ and $\frac{\widetilde{p}%
}{\widetilde{\rho}}\sim\frac{p}{\rho}\sim O(2).$ Moreover, the $5$D
continuous equation of perfect fluid ensures
$\frac{|\partial/\partial t|}{|\partial /\partial
x|}\sim\mathcal{O}(1).$ With all these instruments we can
parametrize any 5D metric theories. Just as in canonical 4D PPN
framework \cite{wilbook}, to get nontrivial results, we should
expand the components of a metric in terms of the linear
combinations of our generalized post Newtonian
potentials to the following orders: $\widetilde{g}_{00}\sim\mathcal{O}%
(4),\widetilde{g}_{0m}\sim\mathcal{O}(3),\widetilde{g}_{mn}\sim\mathcal{O}(2)$.

The concrete relations between the 5D post Newtonian potentials and
the 4D ones can be worked out by means of the Green function. Let
$|\overrightarrow {x}-\overrightarrow{x}^{\prime}|$ be the spatial
distance between the source and field points in the post Newtonian
coordinate system measured by the 4D flat spatial metric, and
$|\vec{x}-\vec{x}^{\prime}|$ be its 3D projection. When
$|\overrightarrow{x}-\overrightarrow{x}^{\prime}|\gg R$, the Green
function $\widetilde{\mathbf{G}}(\overrightarrow{x},\overrightarrow{x}%
^{\prime})$ of the 5D Poisson equation can be approximated as \cite{FL}%
\[
\widetilde{\mathbf{G}}(\overrightarrow{x},\overrightarrow{x}^{\prime}%
)=\frac{G}{|\vec{x}-\vec{x}^{\prime}|}+\frac{2G}{|\vec{x}-\vec{x}^{\prime}|}e^{-\frac{|\vec{x}-\vec{x}^{\prime}|}{R}},
\]
where $G$ is the 4D gravitational constant. Thus we have%
\begin{align*}
\widetilde{U}(\overrightarrow{x}) = \int\widetilde{\mathbf{G}}%
(\overrightarrow{x},\overrightarrow{x}^{\prime})\widetilde{\rho}%
(\overrightarrow{x}^{\prime})dx^{3\prime}dx^{5\prime}=U(\overrightarrow{x})-\gamma\Phi_{2}(\overrightarrow{x})
+\mathcal{O}(6),
\end{align*}
where we used in general $\widetilde{g}_{55} = 1+2\gamma\widetilde
{U}+\mathcal{O}(4)$. By similar ways, we obtain the following
relations:
\begin{align}
\widetilde{\Phi}_{1} &
=\Phi_{1}+\mathcal{O}(6)=\int\frac{G\rho(\vec
{x}^{\prime})v^{2}(\vec{x}^{\prime})}{|\vec{x}-\vec{x}^{\prime}|}
d^{3}x^{\prime}+\mathcal{O}(6),\label{re22}\\
\widetilde{\Phi}_{2} &
=\Phi_{2}+\mathcal{O}(6)=\int\frac{G\rho(\vec
{x}^{\prime})U(\vec{x}^{\prime})}{|\vec{x}-\vec{x}^{\prime}|}d^{3}x^{\prime
}+\mathcal{O}(6),\label{re33}\\
\widetilde{\Phi}_{3} &
=\Phi_{3}+\mathcal{O}(6)=\int\frac{G\rho(\vec
{x}^{\prime})\Pi(\vec{x}^{\prime})}{|\vec{x}-\vec{x}^{\prime}|}d^{3}x^{\prime
}+\mathcal{O}(6),\\
\widetilde{\Phi}_{4} &
=\Phi_{4}+\mathcal{O}(6)=\int\frac{Gp(\vec{x}^{\prime
})}{|\vec{x}-\vec{x}^{\prime}|}d^{3}x^{\prime}+\mathcal{O}(6),\\
\widetilde{V}_{i} &
=V_{i}+\mathcal{O}(5)=\int\frac{G\rho(\vec{x}^{\prime
})v_{i}(\vec{x}^{\prime})}{|\vec{x}-\vec{x}^{\prime}|}d^{3}x^{\prime
}+\mathcal{O}(5),\\
\widetilde{V}_{5} &  =\int\frac{G\rho(\vec{x}^{\prime})\widetilde{v}_{5}%
(\vec{x}^{\prime})}{|\vec{x}-\vec{x}^{\prime}|}d^{3}x^{\prime}+\mathcal{O}(5).
\end{align}

The procedure of parametrizing 5D theories is similar to that of 4D
ones \cite{wilbook}. Here we just outline the main steps and key
points. The
field equation of KK theory with matter fields reads%
\begin{equation}
\widetilde{R}_{\mu\nu}-\frac{1}{2}\widetilde{g}_{\mu\nu}\widetilde{R}%
=8\pi\widetilde{G}\widetilde{T}_{\mu\nu},\label{kkequ}%
\end{equation}
where
$\widetilde{T}_{\mu\nu}=(\widetilde{\rho}+\widetilde{\rho}\widetilde
{\Pi}+\widetilde{p})\widetilde{U}_{\mu}\widetilde{U}_{\nu}+\widetilde
{p}\widetilde{g}_{\mu\nu}$ is the energy-momentum tensor of the 5D
perfect
fluid. Eq.(\ref{kkequ}) is equivalent to%
\begin{equation}
\widetilde{R}_{\mu\nu}=8\pi\widetilde{G}(\widetilde{T}_{\mu\nu}-\frac{1}%
{3}\widetilde{g}_{\mu\nu}\widetilde{T}),\label{kkeq2}
\end{equation}
where
$\widetilde{T}\equiv\widetilde{g}^{\mu\nu}\widetilde{T}_{\mu\nu}.$
The Ricci tensor can be expanded in terms of the perturbative metric
to the
necessary order around the flat background as%
\begin{align}
\widetilde{R}_{00}  & =-\frac{1}{2}\nabla^{2}\widetilde{h}_{00}-\frac{1}%
{2}\sum_{m}\partial_{0}\partial_{0}\widetilde{h}_{mm}+\partial_{m}\partial
_{0}\widetilde{h}_{m0}\nonumber\\
&
+\frac{1}{2}\partial_{m}\widetilde{h}_{00}(\partial
_{n}\widetilde{h}_{mn}-\frac{1}{2}\sum_{n}\partial_{m}\widetilde{h}%
_{nn})
-\frac{1}{4}\nabla\widetilde{h}_{00}\cdot\nabla\widetilde{h}_{00}\nonumber\\
&
+\frac
{1}{2}\widetilde{h}_{mn}\partial_{m}\partial_{n}\widetilde{h}_{00}+\frac{1}%
{2}(\partial_{m}\widetilde{h}_{n0}\partial_{m}\widetilde{h}_{n0}-\partial
_{m}\widetilde{h}_{n0}\partial_{n}\widetilde{h}_{m0}),\label{R00}%
\end{align}%
\begin{align}
\widetilde{R}_{mn} & =
-\frac{1}{2}(\nabla^{2}\widetilde{h}_{mn}-\partial_{m}\partial
_{n}\widetilde{h}_{00}+\sum_{l}\partial_{m}\partial_{n}\widetilde{h}%
_{ll}-\partial_{l}\partial_{n}\widetilde{h}_{ml}\nonumber\\
&
-\partial_{l}\partial
_{m}\widetilde{h}_{nl}),\label{rij}%
\end{align}%
\begin{align}
\widetilde{R}_{0m} =
-\frac{1}{2}(\nabla^{2}\widetilde{h}_{0m}-\partial_{n}\partial
_{m}\widetilde{h}_{0n}+\sum_{n}\partial_{0}\partial_{m}\widetilde{h}%
_{nn}-\partial_{n}\partial_{0}\widetilde{h}_{mn}).\label{r0i}%
\end{align}
The components of the perturbative metric can be solved order by
order.
\begin{itemize}
\item $\widetilde{h}_{00}$ to $\mathcal{O}(2)$: To the required order,
\[
\widetilde{R}_{00}\approx-\frac{1}{2}\nabla^{2}\widetilde{h}_{00},\text{
\ \ }\widetilde{T}_{00}=-\widetilde{T}=\widetilde{\rho},\text{ \ \ }%
\widetilde{g}_{00}=-1.
\]
Thus we have
\[
\nabla^{2}\widetilde{h}_{00}=-\frac{32}{3}\pi\widetilde{G}\widetilde{\rho},\text{
\ \ \ \ } \widetilde{h}_{00} =2\widetilde{U}.
\]
Which justifies that $\widetilde{U}$ is the 5D Newtonian potential.
Note that the $\mathcal{O}(2)$ term of $\widetilde{h}_{00}$ should
be same for any 5D metric theories in order to have the same 5D
Newtonian limitation.

\item $\widetilde{h}_{mn}$ to $\mathcal{O}(2)$: Here we impose the gauge
condition%
\begin{equation}
\frac{1}{2}\partial_{m}\widetilde{h}_{\mu}^{\mu}-\partial_{\mu}\widetilde
{h}_{m}^{\mu}=0.
\end{equation}
From Eq.(\ref{rij}) we have%
\[
\widetilde{R}_{mn}=-\frac{1}{2}\nabla^{2}\widetilde{h}_{mn},
\]
and then
\begin{equation}
-\frac{1}{2}\nabla^{2}\widetilde{h}_{mn}=\frac{8\pi}{3}\widetilde
{G}\widetilde{\rho}\delta_{mn}.\nonumber
\end{equation}
Hence we gets%
\begin{equation}
\widetilde{h}_{mn}=\widetilde{U}\delta_{mn}=U\delta_{mn}.\label{hmn}%
\end{equation}

\item $\widetilde{h}_{0m}$ to $\mathcal{O}(3)$: By imposing the\ gauge
condition%
\begin{equation}
\frac{1}{2}\partial_{0}\widetilde{h}_{\mu}^{\mu}-\partial_{\mu}\widetilde
{h}_{0}^{\mu}=\frac{1}{2}\widetilde{h}_{00,0},\label{gauge}%
\end{equation}
from Eq.(\ref{r0i}) we get
\[
\widetilde{R}_{0m}=-\frac{1}{2}\nabla^{2}\widetilde{h}_{0m},
\]
and thus%
\begin{equation}
-\frac{1}{2}\nabla^{2}\widetilde{h}_{0m}+\widetilde{U}_{,0m}=-8\pi
\widetilde{G}\widetilde{\rho}\widetilde{v}^{m}.
\end{equation}
Hence we obtain%
\begin{equation}
\widetilde{h}_{0i}=-\frac{5}{2}V_{i}-\frac{1}{2}W_{i},\text{ \ }%
h_{05}=3\widetilde{V}_{5},\label{h0m}%
\end{equation}
where
$W_{i}\equiv\int\frac{G\rho(\vec{x}^{\prime})[\vec{v}^{\prime}\cdot
(\vec{x}-\vec{x}^{\prime})](x_{i}-x_{i}^{\prime})}{|\vec{x}-\vec{x}^{\prime
}|^{3}}d^{3}x^{\prime}$ is another 4D post Newtonian potential
\cite{wilbook}.

\item $\widetilde{h}_{00}$ to $\mathcal{O}(4)$: To evaluate this part we use
all the lower-order solutions of $h_{\mu\nu}$. From Eqs.(\ref{R00}),
(\ref{hmn}) and (\ref{h0m}) we get%
\begin{equation}
\widetilde{R}_{00}=-\frac{1}{2}\nabla^{2}\widetilde{h}_{00}-\nabla
^{2}\widetilde{U}^{2}+3\nabla^{2}\widetilde{\Phi}_{2}.
\end{equation}
Thus from Eq.(\ref{kkeq2}) we have
\begin{equation}
\nabla^{2}\widetilde{h}_{00}=2\nabla^{2}\widetilde{U}-2\nabla^{2}\widetilde
{U}^{2}+3\nabla^{2}\widetilde{\Phi}_{1}+2\nabla^{2}\widetilde{\Phi}%
_{2}+2\nabla^{2}\widetilde{\Phi}_{3}+4\nabla^{2}\widetilde{\Phi}_{4},\nonumber
\end{equation}
and hence%
\[
\widetilde{h}_{00}=2\widetilde{U}-2\widetilde{U}^{2}+3\widetilde{\Phi}_{1}+\widetilde{\Phi}_{2}+2\widetilde{\Phi}_{3}
+4\widetilde{\Phi}_{4}.
\]

\end{itemize}

Now we are facing the problem how to relate the parametrized KK
theory to the experiments. For most gravitational experiments in
Solar system, we may consider only the free test particles without
electric charge. From the viewpoint of the KK theory, this implies
that the test particles do not mover along the extra dimension,
i.e., $\widetilde{U}^{\mu}\xi_{\mu}=0$. In this case, it is easy to
show that the 5D geodesic equations for both massive and massless
test particles are reduced to the 4D geodesic equations in the
effective 4D spacetime. Thus the reduced 4D theory behaves just like
a metric theory for these particular test particles or photons.
Along the reduction procedure previously discussed, we can
reduce the parametrized 5-metric to the effective 4-metric $g_{\alpha\beta}$ as%
\begin{align}
g_{00} &  =-1+2U-2U^{2}+3\Phi_{1}+\Phi_{2}+2\Phi_{3}+4\Phi_{4}\label{kk00},\\
g_{0i} &  =-\frac{5}{2}V_{i}-\frac{1}{2}W_{i}\label{kk0i},\\
g_{ij} &  =(1+U)\delta_{ij}.\label{kkij}%
\end{align}
According to the general form of the post Newtonian metric
\cite{wilbook}
\begin{eqnarray}
g_{00}&=&-1+2U-2\beta
U^2-2\xi\Phi_{w}\nonumber\\
&&+(2\gamma+2+\alpha_{3}+\zeta_{1}-2\xi)\Phi_{1}\nonumber\\
&&
+2(3\gamma-2\beta+1+\zeta_{2}+\xi)\Phi_{2}+2(1+\zeta_{3})\Phi_{3}\nonumber\\
&&+2(3\gamma+3\zeta_{4}-2\xi)\Phi_{4}-(\zeta_{1}-2\xi)\mathcal{A}\nonumber\\
&&-(\alpha_{1}-\alpha_{2}-\alpha_{3})w^{2}U-\alpha_{2}w^{i}w^{j}U_{ij} \nonumber\\
&&+(2\alpha_{3}-\alpha_{1})w^{i}V_{i}+\mathcal{O}(6),\\
g_{0i}&=&-\frac{1}{2}(4\gamma+3+\alpha_{1}-\alpha_{2}+\zeta_{1}-2\xi)V_{i}\nonumber\\
&&-\frac{1}{2}(1+\alpha_{2}-\zeta_{1}
+2\xi)W_{i}-\frac{1}{2}(\alpha_1-2\alpha_2)w^{i}U\nonumber\\
&&-\alpha_2 w^j U_{ij}+\mathcal{O}(5),\\
g_{ij}&=&(1+2\gamma U)\delta_{ij}+\mathcal{O}(4),
\end{eqnarray}
the relevant post Newtonian parameters for the reduced KK theory and
4D GR are compared in Table 1.
\begin{table} \caption{The PPN parameters of GR and KK}
\begin{center}
\begin{tabular}{c||c|c|c|c|c}
\hline
\textbf{Theory}&\multicolumn{5}{c}{\textbf{PPN Parameters}}\\
\cline{2-6}& $\gamma$ & $\beta$ & $\xi$ &
($\alpha_1$,$\alpha_2$,$\alpha_3$) & ($\zeta_1$,$\zeta_2$,$\zeta_3$,
$\zeta_4$) \\
\hline
\textbf{GR} & 1 & 0 & 0 & (0,0,0) & (0,0,0,0) \\
\hline
\textbf{KK} & $\frac{1}{2}$ & 0 & 0 & (0,0,0) & (0,0,0,$\frac{1}{6}$)\\
\hline
\end{tabular}
\end{center}
\end{table}

Therefore, given the same (reduced) 4D matter distribution such as a
4D perfect fluid, the detail comparison between the above two
theories leads to significant conclusions. Firstly, the metric
component $g_{00}$ in the post Newtonian coordinates system in KK
theory is smaller than that in 4D GR. But the departure appear only
in $\mathcal{O}(4)$ terms, and hence the reduced 5D KK theory has
the right Newtonian limitation. Secondly, the metric component
$g_{0i}$ in KK theory are $\frac{7}{5}$ times smaller than those in
4D GR in an $\mathcal{O}(3)$ term. This departure may in principle
be detected by the current precise gravitational experiments in
Solar system. At last, the metric components $g_{ij}$ together with
$g_{00}$ and $g_{0i}$ in KK theory determine the post Newtonian
parameter $\gamma = \frac{1}{2}$, which is obviously different from
$\gamma = 1$ in 4D GR. It is obvious that the above departure
is due to the existence of an extra dimension in KK theory.
The disaster of KK theory is that the value
of the parameter $\gamma$ has been accurately measured by Solar
system experiments. In time delay experiment one obtains $\gamma -1=
(2.1\pm2.3)\times10^{-5}$ \cite{r3} \cite{willliving}, and in light
deflection experiment one gets $\gamma -1=(-1.7\pm4.5)\times10^{-4}$
\cite{r1} \cite{r2} \cite{willliving}. Hence there is a severe
contradiction between 5D KK theory and the Solar system experiments.

Our PPN formalism and related discussion can be generalized straightforwardly to
higher dimensional KK theories with compact extra dimensions.
Therefore, although the original idea of Kaluza and Klein is rather
beautiful, the classical KK theories can not survive the
experiments.

\section*{Acknowledgements}

This work is a part of project 10675019 supported by NSFC.


\begin{thebibliography}{99}

\bibitem {kk}T. Appelquist, A. Chodos, and P. G. O. Freund, ed., \emph{Modern
Kaluza-Klein Theories}, (Addison-Wesley, Reading, MA, 1986).

\bibitem {bla}M. Blagojevic, \emph{Gravitation and Gauge Symmetries}, (IOP
Publishing, Bristol, 2002).

\bibitem {kk2}T. Kaluza, Sitz. Preuss. Akad. Wiss. \textbf{33}, 966 (1921).

\bibitem {kk1}O. Klein, Z. Phys. \textbf{37,} 895 (1926).

\bibitem{duff} M.J.~Duff, B.E.W.~Nilsson, and C.N. Pope, Phys. Rep. {\textbf 130}, 1 (1986).

\bibitem {kk3}J. M. Overduin and P. S. Wesson, Phys. Pep. \textbf{283, }305 (1997).

\bibitem {wesson}P.S. Wesson, \emph{Space-Time-Matter: Modern Kaluza-Klein
Theory}, (World Scientific, Singapore, 1999).

\bibitem {str}J. Polchinski, \emph{String Theory I, II}, (Cambridge University Press, 1998).

\bibitem {qq}L. Qiang, Y. Ma, M. Han and D. Yu, Phys. Rev. D \textbf{71,
}061501(R) (2005).

\bibitem {pe}D.J. Gross, M.J. Perry, Nucl. Phys. B \textbf{226}, 29 (1983).

\bibitem {Div}D. Davidson, D. Owen, Phys. Lett. B \textbf{155}, 247 (1985).

\bibitem {we}D. Kalligas, P. S. Wesson and C. W. F. Everitt, Astrophys. J. \textbf{439}, 548 (1995).

\bibitem {we2}P. H. Lim, J. M. Overduin, and P. S. Wesson, J. Math. Phys. \textbf{36}, 6907 (1995).

\bibitem {liu2000}H. Liu, and J. M. Overduin, Astrophys. J. \textbf{538}, 386 (2000).

\bibitem {pe2}R. Mayers and M. Perry, Annal. Phys. \textbf{172}, 304 (1986)

\bibitem {rah}F. Rahaman, S. Ray, M. Kalam and M. Sarker, arXiv: 0707.0951v2.

\bibitem {nor}K.J. Nordtvedt, Astrophys.J. \textbf{161}, 1059 (1970).

\bibitem {will1}K.S. Throne and C.M. Will, Astrophys. J. \textbf{163}, 595 (1971).

\bibitem {will2}C.M. Will, Astrophys. J. \textbf{163}, 611 (1971); \textbf{169}, 125 (1971).

\bibitem {wilbook}C.M. Will, \emph{Theory and experiment in gravitational
physics}, Revised edition (Cambridge University Press, 1993).

\bibitem {willliving}C.M. Will, \emph{The Confrontation between General
Relativity and Experiment}, Living Rev. Relativity \textbf{9}
(2006), 3; http://www.livingreviews.org/lrr-2006-3.

\bibitem {liu}C. D. Hoyle et al, Phys. Rev. D \textbf{70},
042004 (2004).

\bibitem {ma}Y. Ma and J. Wu, Int. J. Mod. Phys. A \textbf{19}, 5043 (2004).

\bibitem {yang}X. Yang, Y. Ma, J. Shao, and W. Zhou, Phys. Rev. D \textbf{68},
024006 (2003).

\bibitem {FL}E.G. Floratos and G.K. Leontaris, Phys. Lett. B \textbf{465}, 95 (1999).

\bibitem {r3}B. Bertotti, L. Iess, and P. Tortora, Nature \textbf{425}, 374 (2003).

\bibitem {r1}S.S. Shapiro, J.L. Davis, D.E. Lebach, and J.S. Gregory, Phys.
Rev. Lett. \textbf{92}, 121101 (2004).

\bibitem {r2}R.N. Treuhaft, and S.T. Lowe, Astron. J. \textbf{102}, 1879 (1991).

\end{thebibliography}
\end{document}